# Green WSUS

## Seifedine Kadry*, Chibli Joumaa

*American University of the Middle East*

**Abstract**

The new era of information and communication technology (ICT) calls for a greater understanding of the environmental impacts of recent technology. With increasing energy cost and growing environmental concerns, green IT is receiving more and more attention. Network and system design play a crucial role in both computing and telecommunication systems. Significant part of this energy cost goes to system update by downloading regularly patches and bug fixes to solve security problems and to assure that the operating system and other systems function properly. This paper describes a new design of Windows Server Update Services (WSUS), system responsible of downloads of the mentioned patches and updates from Microsoft Update website and then distributes them to computers on a network. The general idea behind our proposed design is simple. Instead of the periodical check done by the WSUS servers to ensure update form Microsoft main servers, we rather propose to reverse the scenario in order to reduce energy consumption. In the proposed design, the Microsoft main server(s) sends signal to all WSUS servers to inform them about new updates. Once the signal received, WSUS can contact the main server to start downloading.

## 1. Introduction

Network connectivity has become nearly ubiquitous, and the energy use of the equipment required for this connectivity is growing. Network equipment consists of devices that primarily switch and route Internet Protocol (IP) packets from a source to a destination, edge devices like PCs, servers and other sources and sinks of IP traffic.

Lately, Telecom companies, Internet Service Providers and public organizations reported statistics of network energy requirements and the related pollution -carbon footprint-, showing an alarming and growing trend [1]. The Global e-Sustainability Initiative (GeSI) [2] estimates an overall network energy requirement of about 21.4 TWh in 2010 for European Telcos, and foresees a figure of 35.8 TWh in 2020 if no Green Network Technologies (GNTs) will be adopted. Recently a lot of research is concerned with topics such as green internet or green networking [3]-[4]. The overall goal of these works is to appropriately estimate or measure the energy consumption and cost of the Internet and networking itself. On the other hand they also try to find solutions for decreasing the energy footprint of the Internet and thereby reducing the energy costs. These works most often describe ways to reduce the operating costs of data centers or network devices, such as routers and switches, since these are the most tangible objects and also have the biggest impact. A problem with most of the proposed solutions is that for the given solution the existing protocols on one or more layers and also hardware would have to be modified or completely exchanged. These modifications and replacements would not only be a lot of work which also requires a lot of time until the costs for this restructuring of the system would be amortized. But it also decreases their applicability to the real world tremendously.

---

* Corresponding author. Tel.: +965-66610985.
  *E-mail address*: skadry@gmail.com.

This paper describes an approach which does not require any changes in the underlying protocols. The main idea in this research is to change the static way of windows server update services to be dynamic and energy efficient.

## 2. Why energy savings?

Even though, the energy consumption of the Internet and Internet-related devices is a small amount of the overall energy consumption, it is reasonable to think about a reduction of the energy costs for data centers since their accumulated energy costs has already reached large amounts. Big companies such as Google or Microsoft or also cloud computing providers operate many data centers. For example, the estimated energy costs of Google are approximately $38M, for more information see [5]. If the power consumption could just be reduced by 3% this would lead to energy savings of about a million dollars. It is therefore very desirable to reduce the power consumption for Internet related services, which are usually provided in an accumulated amount by data centers. As the energy consumption of the Internet is rising, as predicted in [6], energy saving approaches for the Internet will be even more profitable. Also growing concern and awareness for environmental protection, green energy usage and sustainability in general are obvious factors which make it even more reasonable to think about alternative concepts in information provisioning.

## 3. WSUS

### 3.1. overview

In Microsoft environment, Operating systems, Office applications and back office applications such as Exchange and SQL Server include millions and millions of lines of code, and it's impossible to predict how the code might be exploited or fail to work in any given situation. Patching is the key to keeping your systems running securely with full functionality. Patching is a hassle. Patching is a problem when you have to get so many systems done and get them done quickly. But patching is so important to maintaining the integrity of the individual systems as well as maintaining the integrity of the entire network. One unpatched system can wreak havoc around the network and bring critical applications to a complete halt. A bigger problem than deploying patches, however, falsely believes that a system is fully and properly patched. This is an area where Microsoft has done a much better job in the last 5 years.
There are many different patch management applications and tools available on the market today. Many of them are targeted at larger installations. When looking at the overall market, Microsoft has positioned three products at three different sizes of organizations [7]:
- Microsoft Update (MU) is targeted toward individual users and small organizations.
- Windows Server Update Services (WSUS) is targeted to mid-sized organizations.
- Systems Management Server is targeted toward enterprise customers.

As previously discussed, the need to perform patching is pretty obvious. With all of the new threats that come out practically every hour, the ability to apply security patches is very important. WSUS enables quick patching where security is vital, and it allows for automation of patching of all systems in the environment. WSUS is a downloadable option for Windows environments that can be used to manage updates for Windows servers and Windows clients on small to large corporate networks.

Our proposed solution is for WSUS, which is the most widely used patching system.

*3.2. Design*

WSUS can be a simple single-server deployment, or it can involve multiple layers of servers. In a basic WSUS environment, as in Fig. 1, the WSUS server on the local area network connects through the Internet to the Microsoft Update server, and all WSUS clients then connect to the WSUS server to get their updates and treat the WSUS server as if it were an MU server.

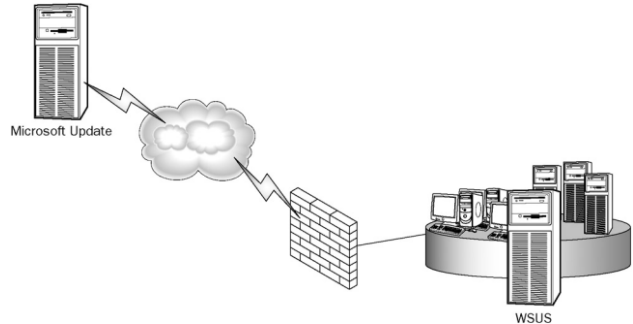

Fig. 1: Basic WSUS environment

Another design scenario includes the use of upstream and downstream WSUS servers. In this type of scenario, as displayed in Fig. 2, the upstream server retrieves all updates from Microsoft Update, and then the downstream WSUS servers get their updates from the upstream server. In this scenario, bandwidth can be better controlled as computers in the downstream network will not consume bandwidth to go to the upstream server.

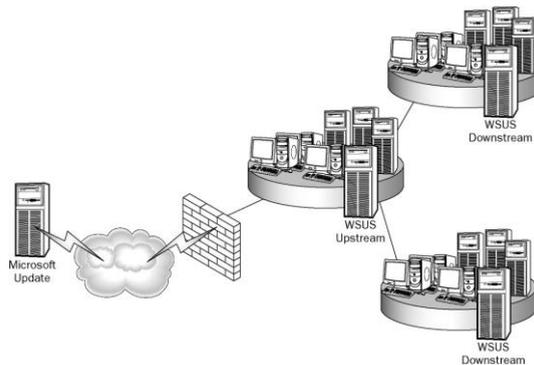

Fig. 2: upstream and downstream design

Microsoft recommends a three-level limit when using upstream and downstream WSUS servers to limit latency as the updates are copied from one level to the next. Using an upstream and downstream WSUS environment, as shown in Fig. 3, allows for greater centralization of WSUS management.

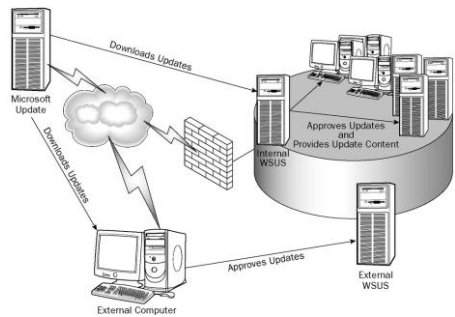

Fig. 3: three level limit

Internal computers can get all of their approvals for updates and the updates themselves from the WSUS server for internal computers. WSUS distributes Microsoft critical updates, definition updates (i.e. for Microsoft Outlook junk email filters and Windows Defender), security updates, update rollups, service packs, and specific tools like the Malicious Software Removal Tool. WSUS contains two types of data, the data about the updates (also called the metadata) and the updates themselves. The metadata is stored in a database so that it can be used in reports and for management of WSUS, while the actual update and patch files themselves are stored in another location.

A WSUS server requests a catalog of updates from either Microsoft's servers, or another WSUS server (aka, an "upstream" server). It posts this catalog to a share and makes it available to clients and downstream WSUS servers to use for assessing which updates they have and do not have. Each client then generates a list of updates it has installed, those that failed to install, and those it needs and sends that list up to the WSUS server. The WSUS administrator (probably you) reviews the cumulative results and approves or declines updates on the WSUS server via the console interface. Clients will then check back on their assigned schedule to identify updates they need which are approved and available on the WSUS server. The clients "pull" these updates and install them and then report back on the results. The clients do most of the heavy lifting. They request update catalogs. They audit themselves for missing updates. They report to the WSUS server as to what they have and what they need. They initiate the download process. They do the installations. They report back their status.

## 4. Green WSUS

As we mentioned previously, the WSUS server or the workgroup user in Windows environment (home or in small business), checks periodically (by default every day at 3:00pm) for updates from Microsoft update (MU), whether an update is available or not, which lead to a waste of traffic and energy. In our proposed design, we let Microsoft update server signals (very small message) a new update to all WSUS servers in order to initiate the download process. The algorithms of the proposed design are:

```
Algorithm: Green MU-WSUS
While (true)
{
If update occurs send broadcast signal to all WSUS servers
}
Algorithm: Green WSUS-MU
For the first time go to Microsoft update, register and synchronize
While (true)
{
If signal received from Microsoft update: start update process
```

```
}
Algorithm: Green PC-WSUS
For the first time go to WSUS local server, register and synchronize
While (true)
{
If signal received from WSUS: start update process
}
```

## 5. Simulation

When a node sends and receives data from other node, the energy consumed is the processing energy (produced by the interface card, the processor, the memory and any other devices that are powered on) and the transmission energy. The following figure (Fig. 4) shows the time evolution and different energy consumed on a node (sender/receiver), where $E_{idle}$ is the energy consumed when the system is idle, $E_{Tx}$ is the transmission energy consumed and $E_{Rx}$ is the reception energy consumed.

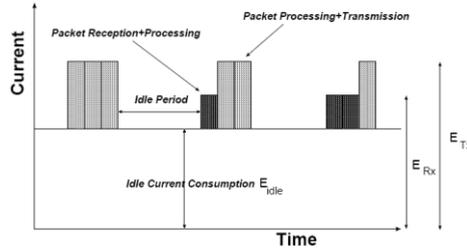

Fig. 4: total energy consumption

The simple expression for the total energy consumed by a node (sender/receiver) is given by:

$E = E_{idle}(t_{total} - t_{Tx} - t_{Rx}) + E_{Tx}t_{Tx} + E_{Rx}t_{Rx}$

Where $E_{idle}$ is the idle energy consumed by the node, $t_{total}$ is the total time needed to complete the transmission process, $t_{Tx}$ and $t_{Rx}$ are the time spent in transmitting and receiving packets, and $E_{Tx}$ and $E_{Rx}$ are the energy consumed at the node for packet transmission and reception. By using the previous expression, two scenarios are simulated:

*Scenario 1*: in this scenario, we suppose that everyday some new updates are available and we evaluate the consumption of energy over one week for a normal update process between Microsoft update and WSUS, and for the green one (proposed design). The results are shown in Fig.5. In this scenario, both designs are very close in terms of energy consumption. However, it should be noted that this scenario is not realistic since Microsoft does not generate daily updates.

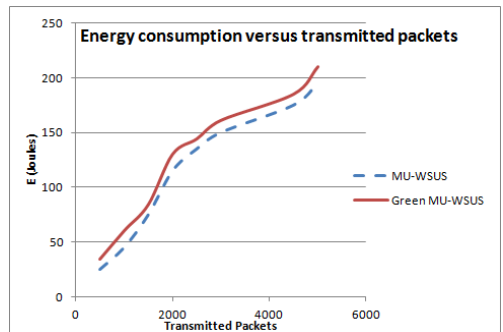

Fig. 5: energy consumption of MU-WSUS v/s Green MU-WSUS

*Scenario 2*: in this scenario, new updates are available once per week. The same comparison as in the previous scenario is done. Fig. 6 shows clearly the reduction of energy consumption using our proposed design in comparison to the typical update procedure.

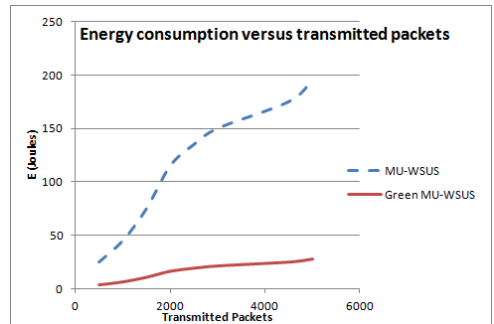

Fig. 6: energy consumption of MU-WSUS v/s Green MU-WSUS

**Conclusion**

In this paper, we proposed a new design to improve the energy consumption in the WSUS server in windows environment, and make the WSUS topology greener. The main idea in the proposed design was to let the Microsoft update server alert all registered WSUS or windows end users about new updates to start the update process. This method has showed a big reduction of the consumed energy when compared to the typical update procedure. This is due to the fact that the proposed method cancels the unneeded checks for updates when none are available.